\documentstyle[12pt,aasms4]{article}

\def\ana{A. \& A.}
\def\mnras{M.N.R.A.S.}
\def\apj{Ap.J.}
\def\apjl{Ap.J. Lett.}
\def\rmp{Revs. Mod. Phys.}

\begin{document}
 
\title{Magnetized Accretion Inside the Marginally Stable Orbit around a
Black Hole}
    
\author{J.H. Krolik\altaffilmark{1}}

\altaffiltext{1}{Department of Physics and Astronomy, Johns Hopkins
University, 3300 N. Charles St., Baltimore, MD 21218; jhk@pha.jhu.edu}
       
\begin{abstract}

     Qualitative arguments are presented to demonstrate that the
energy density of magnetic fields in matter accreting onto a black hole
inside the marginally stable orbit is automatically comparable to
the rest-mass energy density of the accretion flow.  Several consequences
follow: magnetic effects must be dynamically significant, but cannot be so
strong as to dominate; outward energy transport in Alfv\'en waves
may alter the effective efficiency of energy liberation; and vertical
magnetic stresses in this region may contribute to ``coronal" activity. 

\end{abstract}


\section{Introduction}

      Accretion within the last stable orbit around a black hole is
a very complicated problem.  Consequently, most work on the subject
has been carried out in one or another simplifying limit.  Some (as
summarized in the text by Kato et al. 1998) have studied what happens
when the matter behaves
purely hydrodynamically (i.e., without any magnetic effects).  Others,
in order to focus attention on those properties peculiar to MHD,
have taken the limit of ``force-free" magnetic fields, a limit
valid when $B^2/(8\pi) \gg \rho c^2$ (e.g., Blandford \& Znajek 1977,
Okamoto 1992, Ghosh \& Abramowicz 1997).
Still another approach has been to assume an arbitrary magnetic
field configuration, and compute what happens when matter is injected
into the black hole magnetosphere from a specified location at a specified
rate (e.g., Hirotani et al. 1992).  In almost every study, attention has
been restricted to time-steady, azimuthally-symmetric situations
(including, e.g., Phinney 1983 and Punsly 1991, in which the force-free
restriction is relaxed).  One of
the few efforts to go beyond these limits is the simulation by Koide et al.
(1998), which permitted non-stationarity, but imposed azimuthal symmetry,
and used arbitrary initial conditions for both the magnetic field and the
gas pressure distribution.

    In the portion of this literature concentrating on MHD effects,
the key organizing question has been
whether these processes can tap the energy stored in black
hole rotation (Blandford \& Znajek 1977, Phinney 1983, Takahashi et al. 1990,
Punsly 1998, Livio et al. 1998, Meier 1998).

    Unfortunately, these simplifying assumptions, attractive as they are, may
be artificially limiting.  Although one can formally guarantee the validity of
the force-free approximation by setting the accretion rate to zero (as is
sometimes done), real accretion flows with non-zero mass accretion rates may
introduce enough inertia to invalidate that assumption.  Similarly,
if continuing accretion continually brings new plasma toward the
black hole, the field configuration cannot be time-steady.  Furthermore,
simulations of MHD turbulence in the non-relativistic portions of accretion
disks (Stone et al. 1996, Brandenburg et al. 1996) strongly indicate that
the field is far from azimuthally-symmetric.

It is the burden of this {\it Letter}
to argue that the limitations imposed by these simplifying assumptions have
prevented us from seeing a number of important elements in the dynamics of
accretion onto black holes.  Moreover, there has been almost no attention
to the simple question of when these different approximations may apply.
In order to get past these limitations, we
adopt an attitude that is entirely complementary to those
adopted previously.  In order to make the smallest possible number of
limiting assumptions, we will pay the price of relying primarily on
simple scaling arguments, bolstered by one very schematic calculation. 
However, by making the minimum number of assumptions, we will be able to
approach the question of whether the approximations that are commonly
made are likely to be valid.  The question to which we will give the
greatest attention is the range of validity for either the purely hydrodynamic
picture, or the assumption of force-free magnetic field structure.
In addition, rather than
focussing on what happens to the spin of the black hole, we will instead
focus on the fraction of the rest-mass energy of accreting matter that
can be released for possible radiation.

\section{The Magnetic Field Inside the Marginally Stable Orbit}


    We will estimate the state of the magnetic field inside $r_{ms}$,
the radius of the marginally stable orbit, by supposing that it 
is ``frozen" into the accreting plasma.  Its value in the plunging region
can then be scaled from its value in the disk proper by studying the
flow lines.  This view implicitly assumes that the accretion flow
carries negligible net flux, that is, that there are hardly any field
lines extending to infinity.  Otherwise, the field in this region would
be mostly due to the net magnetic flux accumulated over the black hole's
accretion lifetime (as argued, e.g., in Thorne et al. 1986).


    Given the matter four-velocity $u^\alpha$, the field evolution is
conveniently described (Lichnerowicz 1967) by two four-vectors:
$E^\alpha \equiv u_{\beta}F^{\alpha\beta}$ and $B^\alpha \equiv
u_{\beta}(*F)^{\alpha\beta}$,
where $F^{\alpha\beta}$ is the Maxwell field tensor and $*F$ is its dual.
Note that $E^\alpha = (0,\vec E)$ and $B^{\alpha} = (0, \vec B)$ in the fluid
rest frame.  Two of Maxwell's Equations [$\nabla \cdot \vec B = 0$ and
$\nabla \times \vec E = -(1/c)\partial \vec B/\partial t$]
may be written as $(*F)^{\alpha\beta}_{;\beta} = 0$, where we
follow the usual convention and denote a covariant derivative by a semicolon.
The definitions of $E^\alpha$ and $B^\alpha$ may be inverted to write
$*F$ in terms of the field four-vectors, so that this pair of Maxwell's
equations becomes:
\begin{equation}
\left[ u^\alpha B^\beta - u^\beta B^\alpha + 
\epsilon^{\alpha\beta\mu\nu} u_\mu E_\nu \right]_{;\beta} = 0,
\end{equation}
where $\epsilon$ is the completely anti-symmetric Levi-Civita symbol.
Even in these circumstances, the plasma conductivity should easily be
high enough to make $E^\alpha = 0$ everywhere, i.e., to ensure
flux-freezing.  Equation~1 then becomes
\begin{equation}
u^\beta B{^\alpha}_{;\beta} = u^\alpha_{;\beta} B^\beta - 
u^\beta_{;\beta} B^\alpha  - u^\alpha u^\mu u_{\beta ;\mu} B^\beta .
\end{equation}
Here the term $\propto B^{\beta}_{;\beta}$ has been reduced by making use
of the definition of $B^\beta$ and the expression for $*F$ implicitly
given in equation~1.  From equation~2, we see that
the derivative of $B^\alpha$ with respect to proper time comprises
three parts: response to shear (the term $\propto u^\alpha_{;\beta}$);
response to fluid density changes (the term $\propto u^\beta_{;\beta}$);
and response to departures from free-fall
(the term $\propto u^\mu u_{\beta ;\mu}$).

    If only gravitational forces acted on matter with the angular
momentum and energy appropriate to a circular orbit at $r_{ms}$, it would
accelerate inward, but initially rather slowly because both the first
and the second derivatives of the ``effective potential" (in the sense
of Shapiro \& Teukolsky 1983) are zero there.  However,
in a real disk, other forces can also act.   Because the
density of the disk drops sharply inward, there is an inward pressure
gradient force (Chen \& Taam 1993).  In addition, the
fluctuating magnetic fields that remove angular momentum from matter
in the bulk of the disk (Balbus \& Hawley 1998) should work in
essentially the same way here.  It is likely, therefore, that plasma
moves from the nearly-flat portion of the ``effective potential" into
the truly plunging region as a result of non-gravitational forces.
Because the magnetic forces in the disk fluctuate, we expect that the
injection of matter inside $r_{ms}$ will be both time-variabile and
irregular as a function of azimuthal angle.

    However, once in the plunging region proper, one might expect
gravitational forces to again be dominant.  This assumption may
be tested for self-consistency by integrating equations~2 with $u^\alpha$
as given by free-fall with fixed angular momentum $L$ and energy $E$ (as
evaluated at infinity, e.g, Shapiro \& Teukolsky 1983).
Because the metric is time-steady and azimuthally-symmetric, we set
$\partial/\partial t = \partial/\partial \phi = 0$.   For simplicity,
we consider only motion in the equatorial plane, i.e. $u^\theta = 0$,
an idealization appropriate to thin disks.  With regard to the magnetic
field, this calculation may be viewed as describing a particular small
field loop accreted in a way unaffected by any adjacent streams.
Equations~2 writen in Boyer-Lindquist coordinates then reduce to  the single equation
\begin{equation}
u^r {\partial B^{\alpha}_{\rm BL} \over \partial r} = B^{r}_{\rm BL}
{\partial u^\alpha \over \partial r} - B^{\alpha}_{\rm BL} u^{\beta}_{;\beta} .
\end{equation}
The field in the fluid frame is given by
$B^{\alpha}_{\rm fl} = \Lambda^{\alpha}_{\kappa} e^{\kappa}_{\nu}
B^{\nu}_{\rm BL}$,
where the tensor $e^{\kappa}_{\nu}$ gives the orthonormal tetrad for
a locally inertial frame (Chandrasekhar 1983), and
$\Lambda^{\alpha}_{\kappa}$ is the Lorentz
transformation from that frame to the fluid frame.

On the other hand, the rest-mass density of matter $\rho$ as measured in the fluid frame changes according to:
\begin{equation}
{\partial \ln \rho \over \partial r} = - {\partial \ln ( g^{1/2} u^r) \over 
\partial r} ,
\end{equation}
where $-g$ is the determinant of the metric, so that $g^{1/2}$ gives the
scale factor for the differential volume element.  The matter density
can then be expected to fall dramatically as matter crosses $r_{ms}$
because the radial speed outside $r_{ms}$ is
$\simeq \alpha (h/r)^2 v_{orb}$ in a disk with thickness $h \ll r$,
whereas it becomes relativistic inside $r_{ms}$.  Here $\alpha$ is
the usual Shakura-Sunyaev (1973) dimensionless stress, and $v_{orb}$
is the velocity of a circular orbit.

    We numerically integrated equations~3 and 4
along with the equations $dx^\alpha/d\tau = u^\alpha$ (where $\tau$
is the proper time).  To mimic the effect of MHD fluctuations,
we took an initial condition in which $L$ is reduced below the amount
required for a circular orbit at $r_{ms}$ just enough to give the matter an
infall velocity much smaller
than the orbital velocity.  The energy was left fixed at the energy
associated with a circular orbit at $r_{ms}$.

  The result is that the magnetic field in the fluid frame does increase
somewhat inside $r_{ms}$, but not by very large factors.
When $a = 0$, $B_{\rm fl}^r$ is virtually constant until radii
very close to the event horizon, and increases
by only a factor of two even there.  Even if $B_{\rm fl}^\phi$
is zero initially,
the slow growth of the dynamical instability means that the initial orbit
departs only slightly from circular.  Consequently, shear is relatively
strong compared to the change in density, and $B_{\rm fl}^\phi$ can grow to
be comparable to $B_{\rm fl}^r$ in a fraction of an orbit
(if the initial infall velocity is $\ll v_{orb}$).  In fact, the diminution
of $B_{\rm fl}^\phi$ due to falling density dominates its growth due to
shear only if the fluid starts out with an infall speed close to $v_{orb}$.
Otherwise, in the absence of reconnection, $B_{\rm fl}^\phi$ becomes
greater and greater as the matter takes more turns around the black hole,
but diminishes somewhat from its peak in the final approach to the
event horizon.  When $a=0$, the peak $B_{\rm fl}^\phi$ can
be as much as $\simeq 30$ times the initial $B_{\rm fl}^\phi$.
Increasing $a$ from 0 to $0.998M$ (from here on, dimensional expressions
are written in units in which $G = c = 1$) does not change these results 
qualitatively, except that the peak magnitude of $B_{\rm fl}^\phi$
diminishes to $\simeq 1.5$ times the initial value.

At the same time, the dramatic increase in radial speed quickly
leads to a many order of
magnitude decrease in $\rho$.  For example, so long as the initial infall
speed is $\ll v_{orb}(r_{ms})$, when $a=0$ (so that $r_{ms} = 6$),
$u^r = -0.1$ at $r \simeq 4.1$; when $a=0.95$ ($r_{ms} = 1.81$),
the same speed is reached at $r \simeq 1.47$. 

   These results may now be combined to estimate, in the rest frame of
plunging matter, the ratio between the magnetic field energy density
$U_B$ and the rest-mass density $\rho$.  As we have
just shown, $U_B$ is likely to be similar to, or slightly greater than,
the field energy density in the disk.  In terms of the magnetic
contribution to the stress $\alpha_M$, the magnetic energy density in the disk
is $({B_{d}^2 /\langle B_r B_\phi \rangle_d}) \alpha_M p_d $,
where the quantities with subscripts $d$ are measured in the
disk proper.  On the other hand, the rest-mass energy density is
\begin{equation}
\rho  = \left[{g(r_{ms}) \over g(r)}\right]^{1/2}
 \left[{h(r_{ms}) \over r_{ms}}\right]^2
\left[{v_{orb}(r_{ms})  \over u^r (r)}\right]
\left({c \over c_s}\right)^2  \alpha p_d ,
\end{equation}
where $c_s$ is the effective sound speed in the disk.  Because $h/r \simeq c_s/
v_{orb}$, the ratio between magnetic energy density and rest-mass density
in the region of plunging orbits is
\begin{equation}
{U_B \over \rho} \simeq {\alpha_M \over \alpha}
 \left({B_{d}^2 \over \langle B_r B_\phi \rangle_d}\right) 
\left[{B_{\rm fl}^2 (r) \over B_{d}^2}\right]
\left[{g(r) \over g(r_{ms})}\right]^{1/2}
\left[{r_{ms} \over r_{ms}^{3/2} + a/M}\right] u^r (r).
\end{equation}

Numerical MHD simulations of non-relativistic accretion disks
(e.g., Stone et al. 1996, Brandenburg et al. 1996) indicate that $\alpha_M
\simeq \alpha$, and that $B_{d}^2$ is rather greater than $\langle B_r
B_\phi\rangle_d$.  We have just demonstrated that in the free-fall limit,
$B_{\rm fl}^2$ is at least as large as $B_{d}^2$.  The metric
determinant ratio is generally slightly less than unity; if it
is reinterpreted as describing the volume element, the ratio
$[g(r)/g(r_{ms})]^{1/2} = (r/r_{ms}) [h(r)/h(r_{ms}]$ when the infall is
contained within a thin disk centered on the black hole's equatorial
plane.  We conclude, then, that, so long as
magnetic torques play an important role in driving accretion in the
disk proper, {\it the magnetic field must become dynamically important
in the plunging region}.
As a result, it also follows that the assumption of ballistic orbits
is {\it not} self-consistent.


\section{Consequences}

    The simple estimate of equation~6 leads to a number of qualitatively
important consequences.  First, the assumption of purely hydrodynamic
flow is always inappropriate; continuity in $\vec B$ combined
with the dramatic fall in matter density ensures that the magnetic
field is dynamically significant.   Simple free-fall trajectories,
perhaps modified slightly by pressure forces, are
{\it not} a good description of the streamlines; in other words, the
accreting matter gives a significant fraction of its kinetic energy
to the magnetic field.  On the other hand, the force-free
approximation is hardly better except in the extreme limit of zero
accretion and an independently magnetized black hole, for
$B^2/8\pi$ can be no more than comparable
to $\rho$.  This conclusion should be very robust because it
depends only on flux-freezing,
mass conservation, and the assumption that $\alpha_M \simeq \alpha$.
The only way it can be avoided is if there is tremendous flux loss from
the plasma, which would, in its own way, also be very interesting.

    Viewed another way, the estimate of equation~6 shows that the
plunging matter inside $r_{ms}$ must do substantial work on the magnetic
field, transferring much of its kinetic energy to it.  We are then
faced with the question of what happens to this energy.  If it is
simply carried into the black hole along with the plasma, the observational
consequences would be relatively slight.  Two other possibilities are
both plausible and more interesting.

   To the extent that magnetic forces alter the flow, momentum and
energy are carried along field lines.  If field lines
connect matter just outside $r_{ms}$ with matter some ways inside,
the inertia of the plasma outside $r_{ms}$ may cause the field
lines to rotate closer to its angular frequency than the considerably
larger angular frequency of matter with the same angular momentum that
has fallen to smaller radius.  A torque would then be exerted in which
angular momentum is carried back to the disk proper.  The associated
work done on the disk would ultimately be dissipated, effectively
increasing the radiative efficiency of the disk.    
That the field likely takes the form of stretched loops does not diminish
this effect:
the sense of orbital shear automatically gives $B^r B^\phi$ the same sign
no matter what the sign of $B^r$ is.  Note that insisting on azimuthal
symmetry would eliminate this sort of field structure.

  How much energy and angular momentum is carried outward depends critically
on, among other things, whether the MHD waves travel rapidly enough that
waves directed outward in the fluid frame actually move outward in the
coordinate frame.  It is at this point in the argument that the
problem of where causal disconnection sets in becomes central.
To quantify this issue, suppose that the dispersion
relation for linear waves with wave-vector $\vec k = + k\hat r$
in the fluid frame is $\omega^2 = v^2 k^2$, so that the phase and
group speeds coincide.  For the waves to move outward in the coordinate
frame, $\omega$ must stay positive when the wave four-momentum is
transformed to that frame.  By this definition, if the matter follows
ballistic trajectories, the minimum wave
speed in the fluid frame $v_{min} \simeq |u^r|$ when $a = 0$;
when $a > 0$, the ratio $v_{min}/|u^r|$ increases with both increasing
$a$ and decreasing $r$, and can be as large as $\sim 10$ when $a = 0.998$.
Since $B_{\rm fl}^2 \sim 8\pi \rho$, we expect the Alfv\'en speed
in the fluid frame to be $\sim c$, so it may be that $v > v_{min}$
in much of the inflow region, particularly when $a$ is relatively small.
However, a proper evaluation of this
mechanism's efficiency clearly requires a much more complete
calculation, and that is far beyond the scope of this paper.

    The speculation that Alfv\'en waves may carry away significant energy
seems to be at variance with the long-held belief
that the maximum energy available for radiation is the binding energy of
matter at the marginally stable orbit.  The traditional argument has been
that so little time is required for matter to fall from $r_{ms}$ to
the vicinity of the event horizon that the relatively slow processes of kinetic
dissipation have no opportunity to transform any more of its energy into
heat, and thence into photons.  However, there is a loophole in this
argument---Alfv\'en wave radiation is a coherent process that may transfer
a substantial amount of energy on the dynamical time.

    The assumption of time-steadiness becomes especially limiting in this
regard.  If one requires the field structure to be time-steady,
there can be no travelling waves.  However, when the magnetic field inside
$r_{ms}$ remains connected to the disk (as it must, at least initially, if
it is carried in by accreting matter), it cannot be time-steady except in
some long-time average sense. 

     If significant energy can be removed from plunging matter, carried
back out to the disk, and dissipated just outside $r_{ms}$, the
effective efficiency of accretion may be greater than that given by the
binding energy at $r_{ms}$.  In a Schwarzschild metric, in principle one
might imagine that the efficiency might approach unity; in a Kerr metric,
it is even imaginable that the outward flux of angular momentum carried by
the magnetic field is so strong that the matter between $r_{ms}$ and
the event horizon is put on negative energy orbits.  If that occurs, the
efficiency would become nominally {\it greater} than unity because the
accumulated spin energy of the black hole is being tapped.  In other words,
there is the potential here for a realization of the Penrose process, made
feasible by the long-range action of magnetic fields (Phinney 1983,
Okamoto 1992, Yokosawa 1993).
Both upper bounds are, however, somewhat unlikely to be realized
in practise.  Efficient removal of energy would mean that, at any given
radial coordinate, the kinetic energy of the flow would be smaller than
in free-fall.  If this occurs through a reduction in $|u^r|$, the relative
importance of magnetic forces would be diminished, perhaps leading to
a self-limiting of this process.  Similarly, the ``Penrose process"
would likely require the orbital frequency of the matter inside
$r_{ms}$ to be reduced below the orbital frequency of the matter in the
disk to which it is donating its angular momentum.

     Whether Alfv\'en waves can carry a significant amount of energy and
angular momentum outwards to a ``normal" accretion disk is also an issue
in the theory of advection-dominated accretion flows (ADAFs).  As described,
for example, by Narayan \& Yi (1995), these are accretion flows in which
the inflow speed is close to freefall, and in which angular momentum
is efficiently removed to larger radii by magnetic stresses.  Their mechanics
are therefore very similar to the mechanics of accretion inside $r_{ms}$
as envisaged here.  The difference here is that there is ``normal" disk
at radii only factors of a few outside the region of interest, where
the energy carried outward can be dissipated.  In contrast, most ADAFs
are supposed to span a large range in radius.

   ``Coronal" activity is a second possible outlet for the magnetic
energy.  If
$B^2 \sim 8\pi \rho$ in the accreting matter, but is weaker for $|z| > h$,
the vertical component of the $\nabla B^2$ force would be $\sim r/h$ times
the vertical gravity of the black hole.  Strong upwelling of matter must
then result, leading directly to the creation of a significant vertical
component in the magnetic field.
Although these vertical motions are, like all the motions in this region,
relativistic, flux loss into the corona cannot reduce the field strength
by more than factors of a few because the inward plasma flow amplifying
the field is equally rapid.

     As magnetic loops rise vertically, the shearing of their foot points
can be expected to lead to reconnection in exactly the same fashion as
expected in the non-relativistic portions of accretion disks (e.g.,
Galeev et al. 1979, Romanova et al. 1998).  So much energy is available
that the associated dissipation could be
a major contributor to the ``coronal" activity (i.e., strong hard X-ray
emission) seen so often in accreting black hole systems.  Because Compton
scattering is such an efficient energy-loss mechanism, the dissipated heat
could be radiated in less than a dynamical time:
\begin{equation}
t_{\rm Compt} \sim \left({m_e \over \mu_e}\right)
\left({L \over L_E}\right)^{-1},
\end{equation}
where $\mu_e$ is the mean mass per electron and $L/L_E$ is the luminosity of
seed photons in terms of the Eddington luminosity.
In a normal plasma, $m_e/\mu_e \sim 10^{-3}$, of course.

     A further consequence of the strong vertical forces that may be
expected is the possible excitation
of jets.  This mechanism appears to provide an efficient way both to eject
magnetized plasma along the rotation axis of the system, and to heat it
at the same time.

    In sum, many of the simplifying assumptions on which intuition about
accretion inside the marginally stable orbit has been built may have misled
us: Supposing the flow to be time-steady prevents recognition of both
travelling waves and unsteady behavior like magnetic reconnection; supposing
it to be azimuthally-symmetric forbids consideration of magnetic field
topologies with stretched loops; supposing it to be purely hydrodynamic
eliminates the possibility of magnetically-mediated dynamics; and supposing
that the magnetic field is force-free rules out consideration of how
infall dynamics might alter the field configuration.

\acknowledgments

     I am happy to acknowledge stimulating and instructive
conversations with Eric Agol,
Mitch Begelman, Roger Blandford, Jim Pringle, and Ethan Vishniac.  This
work was partially supported by NSF Grant AST-9616922 and NASA
Grant NAG5-3929.


\begin{references}

\reference{bh98} Balbus, S.A. \& Hawley, J.F. 1998, \rmp\  70, 1
\reference{bz77} Blandford, R.D. \& Znajek, R.L. 1977, \mnras\  179, 433
\reference{b96} Brandenburg, A., Nordlund, A., Stein, R.F., \& Torkelsson, U.
1996, \apjl\ 458, L45
\reference{c83} Chandrasekhar, S. 1983, The Mathematical Theory of Black Holes
(Oxford: Oxford University Press)
\reference{ct93} Chen, X.-M. \& Taam, R.E. 1993, \apj\  412, 254
\reference{grt79} Galeev, A.A., Rosner, R. \& Vaiana, G. 1979, \apj\ 229, 318
\reference{ga97} Ghosh, P. \& Abramowicz, M. 1997, \mnras\ 292, 887
\reference{htnt92} Hirotani, K., Takahashi, M., Nitta, S. \& Tomimatsu,
A. 1992, \apj\  386, 455
\reference{kfm98} Kato, S., Fukue, J. \& Mineshige, S. 1998, Black-Hole
Accretion Disks (Kyoto: Kyoto University Press)
\reference{k98} Koide, S., Shibata, K. \& Kudoh, T. 1998, \apjl\ 495, L63
\reference{l67} Lichnerowicz, A. 1967, Relativistic Hydrodynamics and
Magnetohydrodynamics (New York: Benjamin)
\reference{lop98} Livio, M., Ogilvie, G. \& Pringle, J.E. 1998, preprint
astro-ph/9809093
\reference{m98} Meier, D. 1998, preprint astro-ph/9810352
\reference{ny95} Narayan, R. \& Yi, I. 1995, \apj\  452, 710
\reference{o92} Okamoto, I. 1992, \mnras\ 254, 192
\reference{p83} Phinney, E.S. 1983, unpublished Cambridge University Ph.D.
thesis
\reference{p91} Punsly, B. 1991, \apj\  372, 424
\reference{p98} Punsly, B. 1998, \apj\  506, 790
\reference{r98} Romanova, M.M., Ustyugova, G.V., Koldsba, A.V.,
Chechetkin, V.M. \& Lovelace, R.V.E. 1998, \apj\ 500, 703
\reference{ss73} Shakura, N.I. \& Sunyaev, R.A. 1973, \ana 24, 337
\reference{st83} Shapiro, S.L. \& Teukolsky, S.A. 1983, Black Holes, White
Dwarfs, and Neutron Stars (New York: Wiley-Interscience)
\reference{s96} Stone, J.M., Hawley, J.F., Gammie, C.F., \& Balbus, S.A.
1996, \apj\  463, 656
\reference{t90} Takahashi, M., Nitta, S., Tatematsu, Y. \& Tomimatsu, A.
1990, \apj\  363, 206
\reference{tpm86} Thorne, K.S., Price, R.H. \& Macdonald, D.A. 1986, Black
Holes: The Membrane Paradigm (New Haven: Yale University Press)
\reference{y93} Yokosawa, M. 1993, P.A.S.J. 45, 207
\end{references}
\end{document}